\documentclass[prd, aps, preprint,nofootinbib,superscriptaddress,12pt]{revtex4-1}
\pdfoutput=1

\usepackage{graphicx}
\usepackage{soul}
\usepackage{amsmath}
\usepackage[makeroom]{cancel}
\usepackage{empheq}

\usepackage{subfigure}
\usepackage{amssymb}

\usepackage{enumitem}

\usepackage{graphics}

\usepackage{bbold}

\usepackage{hyperref}

\usepackage[compat=1.0.0]{tikz-feynman}

\usepackage{physics}

\newcommand{\nn}{\nonumber}
\renewcommand{\textit}[1]{#1}

\newcommand{\aKD}{a_{CP}^{K,D}}
\newcommand{\aKK}{a_{CP}^{K,\epsilon}}
\newcommand{\Kbar}{\overline{K}}

\usepackage{slashed}

\usepackage[english]{babel}
\usepackage{amsthm}

\def\beq{\begin{equation}}
\def\eeq{\end{equation}}
\def\beqa{\begin{eqnarrbay}}
\def\eeqa{\end{eqnarray}}
\def\ben{\begin{enumerate}}
\def\een{\end{enumerate}}

\usepackage{import}
\usepackage{slashed}

\begin{document}

\author{Yuval~Grossman}
\email{yg73@cornell.edu}
\affiliation{LEPP, Department of Physics, Cornell University, Ithaca, NY 14853, USA}

\author{Guglielmo~Papiri}
\email{gp343@cornell.edu}
\affiliation{LEPP, Department of Physics, Cornell University, Ithaca, NY 14853, USA}

\author{Stefan~Schacht}
\email{stefan.schacht@durham.ac.uk}
\affiliation{Institute for Particle Physics Phenomenology, Department of Physics,
Durham University, Durham DH1 3LE, United Kingdom}

\preprint{IPPP/26/34}

\title{$CP$ violation in neutral kaon mixing in $D^0\rightarrow K_SK_S$}

\begin{abstract}
We study CP violation induced by neutral kaon mixing in $a_{CP}(D^0 \rightarrow K_S K_S)$. We show that the contribution from neutral kaon mixing arises only in connection with second-order weak interactions in $D$ decays. We estimate this effect to be at the $10^{-6}$ level, and thus negligible compared to current experimental sensitivity and to the expected contribution from CP violation in the charm sector. Similar reasoning can be applied to other systems.
\end{abstract}

\maketitle

\section{Introduction}

The $D^0 \to K_S K_S$ decay is interesting in several aspects. 
It is a rare decay \cite{ParticleDataGroup:2024cfk},
\begin{equation}
{\rm BR}(D^0 \to  K_S K_S) = (1.41\pm 0.05) \times 10^{-4},
\end{equation}
as it is suppressed by several factors:
\begin{enumerate}
\item It is a singly Cabibbo-suppressed (SCS) process.
\item It involves the spectator quark.
\item The CKM-leading amplitude vanishes in the U-spin limit~\cite{Savage:1991wu}.
\end{enumerate}
The CP asymmetry,
\begin{equation}\label{eq:CP_intro}
a_{CP}(D^0\rightarrow K_S K_S) \equiv \frac{\Gamma(D^0\rightarrow K_S K_S)-\Gamma(\overline{D}^0\rightarrow K_S K_S)}{\Gamma(D^0\rightarrow K_S K_S)+\Gamma(\overline{D}^0\rightarrow K_S K_S)},
\end{equation}
is particularly interesting. The third suppression factor mentioned above, namely the U-spin argument, implies that the CP asymmetry is enhanced by a factor of the order of the inverse of the dimensionless U-spin breaking parameter, typically assumed to be $\mathcal{O}(20\%)$, resulting in a prediction that the asymmetry is expected to be of the order of a few permill, and can be as high as $1\%$ within the SM~\cite{Hiller:2012xm,Nierste:2015zra}. 
This enhancement makes this decay particularly interesting for probing CP violation in charm. Experimentally, the time-integrated CP asymmetry in this mode has been measured by 
CLEO \cite{CLEO:2000exp}, 
Belle and Belle~II \cite{Belle:2017exp, Belle2:2024exp, Belle:2025cub},
LHCb~\cite{LHCb:2015ope,LHCb:2018exp,LHCb:2021exp,LHCb:2025exp} 
and 
CMS~\cite{CMS:2024exp}. 
The current world average is~\cite{Tuci:2025}
\begin{align}
a_{CP}(D^0\rightarrow K_S K_S ) &= (-0.17 \pm 0.62 \pm 0.18)\%\,.
\end{align}
Further improvements are expected as LHCb and Belle~II accumulate larger  datasets~\cite{Belle-II:2018_physics_book,LHCb:2018_new_upgrades}.

In this work we investigate the contribution to the CP asymmetry, Eq.~(\ref{eq:CP_intro}), due to neutral kaon mixing and oscillations in the final state. 
CP violation in neutral kaon mixing has been studied in the literature for decays with one neutral kaon in the final state \cite{Bigi:2005ts,Xing:1995jg,Yu:2017oky,Grossman:2011,Grossman:2025}. It is known that modes such as $D^{\pm}\rightarrow \pi^{\pm} K_S$, $D^{\pm}\rightarrow K^{\pm}K_S$, and $\tau^{\pm}\rightarrow \pi^{\pm}K_S\nu$ exhibit CP asymmetries that come from kaon mixing at the order of $2\, \mathrm{Re}(\epsilon)\sim 3 \times 10^{-3}$, where $\epsilon$ is the kaon CP violating parameter. 
In Ref.~\cite{Grossman:2011} it was shown  that the measured CP asymmetry due to neutral kaon mixing depends on the efficiency function of the experiment to reconstruct $K\rightarrow\pi\pi$ as a function of the rest-frame decay time of the neutral kaon. 

The case where there are two neutral kaons, $K^0$ and $\overline{K}^0$, in the final state was recently studied 
in Ref.~\cite{Grossman:2025}.
Without taking into account the efficiency function, the kaon mixing effect cancels, as the effect from the final $K^0$ cancels the effect from the $\overline{K}^0$. Yet, once the  efficiency function is taken into account, that cancellation is not perfect and it leads to a kaon induced  CP asymmetry for decays with two neutral kaons in the final state. The goal of this work is to study this effect in the case of $D^0 \to K_S K_S$, which is a mode with two correlated neutral kaons in the final state.

In this paper we present three main points:
\begin{itemize}
    \item We discuss the formalism to compute the CP asymmetry due to neutral kaon mixing in decays where two correlated neutral kaons are detected.
    \item We show that in the Standard Model (SM), at leading order in the weak interactions, there is no CP violation in kaon mixing for decays into a final state with two correlated neutral kaons, even including the efficiency effects discussed in Ref.~\cite{Grossman:2025}.
    \item We show that CP violation in neutral kaon mixing in $D^0\rightarrow K_SK_S$ arises from second order weak interactions. We estimate this effect and find that the result is negligible for realistic experimental precision. 
\end{itemize}

\section{Definitions and experimental parameters}

We follow the same formalism as in Ref.~\cite{Grossman:2025}. There are three sources for the experimentally measured CP asymmetry. Assuming that all of them are small, these contributions can be treated additively
\begin{equation}
\label{eq:sep-acp}
    a_{CP,exp} = a_{CP}^{D} + a_{CP}^K + a_{CP}^{\text{det}}\;,
\end{equation}
where $a_{CP}^{D}$ is the fundamental CP asymmetry in the $D^0\rightarrow K^0 \overline{K}^0$ decay (which comes from the direct CP violation in the meson decay, and from the indirect CP violation in $D^0$ and $\overline{D}^0$ mixing); $a_{CP}^K$ is the CP asymmetry in the kaon system, due to neutral kaon mixing; $a_{CP}^{\text{det}}$ is the detector asymmetry due to different detection efficiencies of tagging the flavor of the $D$ meson. 

In the following we only focus on $a_{CP}^K$, and we show that it separates into two different effects:
\begin{equation}\label{eq:aKCP_separation}
    a_{CP}^K = \aKK + \aKD\;.
\end{equation}
$\aKK$ is the asymmetry due to CP violation in kaon oscillations in the final state. $\aKD$ is the asymmetry due to CP violation in the $D$ decay, that requires kaon mixing in order to be probed. $\aKK$ requires CP violation in the kaon system, while $\aKD$ contributes even in the $\epsilon = 0$ limit.

We use the standard formalism: 
\begin{align}\label{eq:pq}
    |K_S\rangle = p\,|K^0\rangle + q\,|\overline{K}^0\rangle\;, \qquad
    |K_L\rangle = p\,|K^0\rangle - q\,|\overline{K}^0\rangle\;,
\end{align}
where $\ket{K_{L,S}}$ are the kaon mass eigenstates, and we have~\cite{ParticleDataGroup:2024cfk}
\begin{equation}
    \frac{|p|^2-|q|^2}{|p|^2+|q|^2}\approx 2\, \text{Re}(\epsilon) =  (3.192\pm 0.026)\times10^{-3}\;,
\end{equation}
where $\epsilon$ is the kaon CP violation parameter. Moreover, we use the standard definitions:
\begin{align}
    \Delta m = m_L - m_S\;,\qquad
    \Delta \Gamma = \Gamma_L - \Gamma _S\;,\qquad \Gamma=\frac{\Gamma_S + \Gamma_L}{2}\;.
\end{align}
We also define $A_{S,L}$ to be the decay amplitudes of the kaon mass eigenstates into two pions:
\begin{equation}
A_{S,L}=\bra{\pi\pi}H\ket{K_{S,L}}\;,\qquad \frac{A_L}{A_S}\approx\epsilon\;.
\end{equation}

We emphasize that in the $D^0\rightarrow K_S K_S$ decay the $K_S$ in the final state are experimentally identified as a system of two pions with invariant mass $m_{\pi\pi}\approx m_K$ (we neglect effects of order $\epsilon'$ and thus we do not distinguish between $\pi^+\pi^-$ and $\pi^0\pi^0$). While this is an excellent approximation to a true $K_S$, it is not identical. This difference can have important implications when considering CP asymmetries at the level of $10^{-3}$, see Refs.~\cite{Grossman:2011,Grossman:2025}.

\section{Contributions at second order of the weak interactions}\label{sec:second_order}

In terms of kaon interaction eigenstates, there are three relevant amplitudes which contribute to the decay channel $D^0\rightarrow K_S K_S$:
\begin{align}\label{eq:Amplitueds}
\mathcal{A}_{K\overline{K}} &= \mathcal{A}(D^0\rightarrow K^0\overline{K}^0)\;,\nonumber\\
\mathcal{A}_{KK} &= \mathcal{A}(D^0\rightarrow K^0 K^0)\;,\\
\mathcal{A}_{\overline{K}\overline{K}} &= \mathcal{A}(D^0\rightarrow \overline{K}^0\overline{K}^0)\;.\nonumber
\end{align}
We also define the ratios:
\begin{align}\label{rKK}
    r_{\overline{K}\overline{K}} &= \frac{\mathcal{A}(D^0 \to \overline{K}^0\overline{K}^0)}{\mathcal{A}(D^0 \to K^0\overline{K}^0)}\;,\qquad r_{\overline{K}\overline{K}}=|r_{\overline{K}\overline{K}}|e^{i(\phi_r+\delta_r)}\;,\\
    r_{KK} &= \frac{\mathcal{A}(D^0 \to K^0K^0)}{\mathcal{A}(D^0 \to K^0\overline{K}^0)}\;, \qquad 
r_{KK}=|r_{KK}|e^{i(\phi_{s}+\delta_s)}\;.
\end{align}
Here, $\delta_{r,s}$ are the relative strong phases, and $\phi_{r,s}$ are the relative weak phases between the amplitudes.
Experimentally, a system of four pions in the final state $D^0\rightarrow K_S(\rightarrow \pi\pi)K_S(\rightarrow\pi\pi)$ is detected. Thus, all three amplitudes correspond to indistinguishable processes that need to be summed coherently. Note that within the SM, $\mathcal{A}_{K\overline{K}}$ is the dominant contribution, as $\mathcal{A}_{KK}$ and $\mathcal{A}_{\overline{K}\overline{K}}$ are $|\Delta S|=2$ processes which in the SM  only occur at second order in the weak interactions. We also note that the CKM-leading diagrams that contribute to $\mathcal{A}_{K\overline{K}}$, which are proportional to the singly Cabibbo suppressed factor $\lambda_{sd}=(V_{cs}^*V_{us}^{\phantom{*}} - V_{cd}^*V_{ud}^{\phantom{*}})/2$, vanish in the limit of exact U-spin symmetry~\cite{Savage:1991wu}. This suppression does not apply to the $|\Delta S|=2$ amplitudes, and therefore we estimate that the ratios $r_{KK}$ and $r_{\overline{K}\overline{K}}$ are enhanced by a factor $1/\varepsilon_{\text{U-spin}}$. Here, $\varepsilon_{\text{U-spin}}$, assumed to be of $\mathcal{O}(20\%)$, denotes the U-spin breaking parameter, whose size is empirically motivated by the ratio of decay constants of the kaon and pion, which is determined by Lattice QCD as $f_K/f_\pi-1\sim 0.2$ \cite{FlavourLatticeAveragingGroupFLAG:2024oxs, Bazavov:2017lyh, Dowdall:2013rya, Carrasco:2014poa, Miller:2020xhy, ExtendedTwistedMass:2021qui}.

\begin{figure}[t]
\subfigure[\, $E_{\overline{K} \overline{K}}$]{\includegraphics[width=0.24\textwidth]{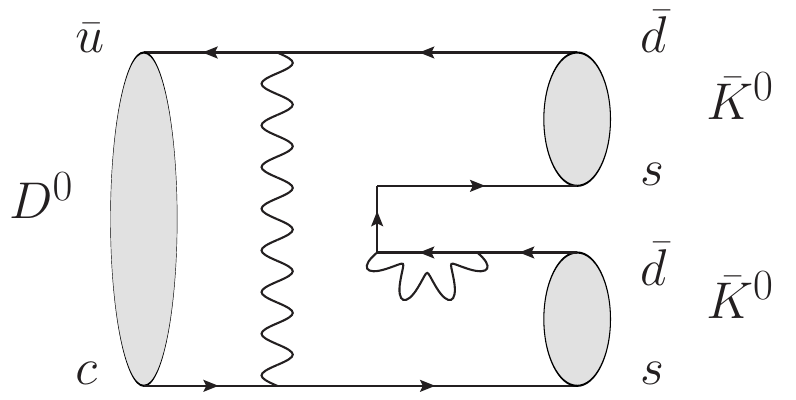}}
\subfigure[\, $C_{\overline{K} \overline{K}}$]{\includegraphics[width=0.24\textwidth]{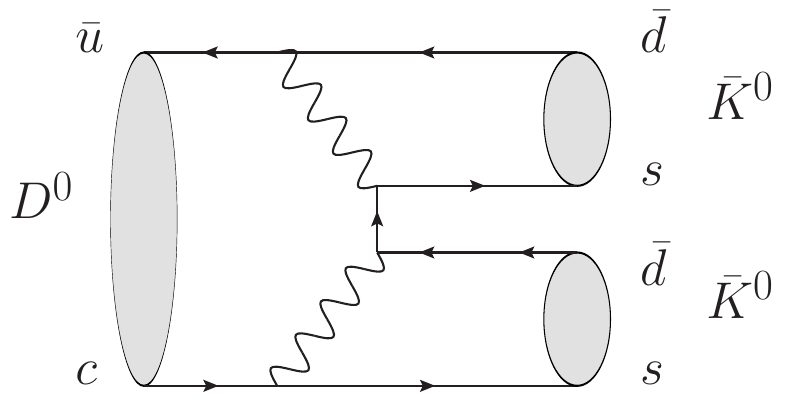}}
\subfigure[\, $E_{KK}$]{\includegraphics[width=0.24\textwidth]{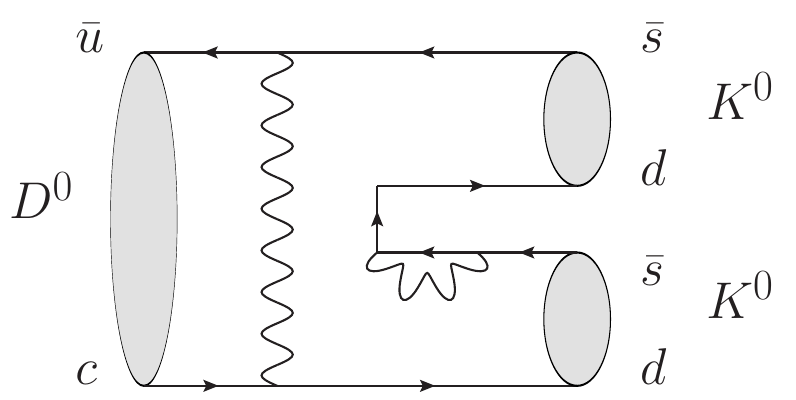}}
\subfigure[\, $C_{KK}$]{\includegraphics[width=0.24\textwidth]{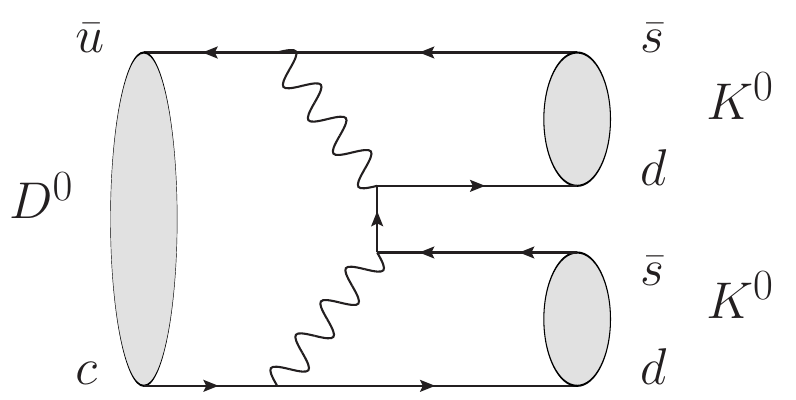}}
    \caption{
       Diagrams which contribute to the second order weak interactions for the $D^0\rightarrow \overline{K}^0\overline{K}^0$ decay ($E_{\overline{K}\overline{K}}$ and $C_{\overline{K}\overline{K}}$) and for the $D^0\rightarrow K^0K^0$ decay ($E_{KK}$ and $C_{KK}$). Feynman diagrams created with JaxoDraw~\cite{Binosi:2003yf, Binosi:2008ig}.
    }
    \label{fig:diagrams}
\end{figure}

In this section we estimate the size of the parameters $r_{\overline{K}\overline{K}}$ and $r_{KK}$. In Fig.~\ref{fig:diagrams} we show the diagrams, denoted as $C_{\overline{K}\overline{K}},E_{\overline{K}\overline{K}}$ and $C_{KK},E_{KK}$, that contribute to the amplitude for $D^0\rightarrow \overline{K}^0\overline{K}^0$ and $D^0\rightarrow K^0K^0$, respectively. We further define  
\begin{equation}
r_{\overline{K}\overline{K}} = r_{\overline{K}\overline{K}}^E + r_{\overline{K}\overline{K}}^C\;,\quad r_{KK}=r_{KK}^E + r_{KK}^C\;.
\end{equation}

We start with an order of magnitude estimate of the loop diagram $E_{\overline{K}\overline{K}}$. Using the unitarity of the CKM matrix, we find:
\begin{align}\label{eq:rKbarKbarE}
    r_{\overline{K}\overline{K}}^E=\frac{E_{\overline{K}\overline{K}}}{\mathcal{A}_{K\overline{K}}} &\sim \frac{G_Fm_D^2}{16\pi^2\;\varepsilon_{\text{U-spin}}}\times\frac{V_{cs}^{*}V_{ud}^{\phantom{*}} 
    \left[ V_{ts}^* V_{td}^{\phantom{*}} (f(x_t)-f(x_u)) + 
        V_{cs}^* V_{cd}^{\phantom{*}} (f(x_c)-f(x_u))  \right]}{\lambda_{sd}} \\
    %%%
    &\sim \frac{G_Fm_D^2}{\varepsilon_{\text{U-spin}}}\times\frac{\lambda^4 x_t}{16\pi^2}\sim 10^{-8}\;, \nn
\end{align}
where $f(x_q)$ is a loop function depending on $x_q=m_q^2/m_W^2$, $G_Fm_D^2$ comes from the second order weak interaction suppression, $1/16\pi^2$ comes from the relative loop suppression, $1/\varepsilon_{\text{U-spin}}$ is the enhancement due to the U-spin symmetry suppression of $\mathcal{A}_{K\overline{K}}$, and $\lambda\approx0.225$ is the CKM Wolfenstein parameter.
For our rough estimate, we use $f(x_q)\sim x_q$ in order to capture the effect of the GIM mechanism. The interplay of the latter with the CKM suppression results in
\begin{align}
\frac{V_{ts}^* V_{td}^{\phantom{*}}}{V_{cs}^* V_{cd}^{\phantom{*}}} \frac{f(x_t)}{f(x_c)} \sim \lambda^4\frac{x_t}{x_c} \sim 50\,. \label{eq:CKM-GIM-interplay}
\end{align}
We note that Eq.~(\ref{eq:CKM-GIM-interplay}) is a reflection of hard GIM suppression as opposed to a soft GIM mechanism with a logarithmic dependence on the quark masses. The latter would occur in case of photon penguin contributions, see Refs.~\cite{Buchalla:1995vs, Buchalla:2001ux}.

Next we estimate the size of $C_{\overline{K}\overline{K}}$. We note that this is a tree-level diagram, and we find:
\begin{align}\label{eq:C_KbarKbar}
    r_{\overline{K}\overline{K}}^C=\frac{C_{\overline{K}\overline{K}}}{\mathcal{A}_{K\overline{K}}} &\sim G_F m_D^2\times \frac{(V_{cs}^*V_{ud}^{\phantom{*}} V_{us}^*V_{ud}^{\phantom{*}})}{\lambda_{sd}}\times\frac{1}{\varepsilon_{\text{U-spin}}} 
    %%%
    \sim \frac{G_F m_D^2}{\varepsilon_{\text{U-spin}}}\sim 10^{-4}\;.
\end{align}
We note that while the topology $C_{\overline{K}\overline{K}}$ and the leading exchange topology of $D\rightarrow K^0 \overline{K}^0$  are very different, we do not expect large long-distance corrections to our estimate of $C_{\overline{K}\overline{K}}$, because, as a consequence of the $\Delta S=2$ transition, no QCD rescattering of the final state quarks ($\bar{d} \bar{d}$) and ($\bar{s} \bar{s}$) is possible. Such effects would be suppressed by even another order in the weak interaction. 

Note further that for the estimate in Eq.~(\ref{eq:C_KbarKbar}), we write the contribution with an $u$ quark in the internal propagator. 
Since the energy scale of the diagram is of order $m_D$, at the leading order in the Wolfenstein CKM parametrization the charm quark contribution only modifies this estimate by an $\mathcal{O}(1)$ correction, and the top quark contribution is suppressed by a relative correction $\mathcal{O}(\lambda^{5}\, m_D/{m_t})\sim 10^{-5}$.

Analogously, for the diagrams which contribute to the amplitude for $D^0\rightarrow K^0K^0$ we find, see Fig.~\ref{fig:diagrams}:
\begin{align}
    r_{KK}^E=\frac{E_{KK}}{\mathcal{A}_{K\overline{K}}}&\sim \frac{G_F m_D^2}{16\pi^2\;\varepsilon_{\text{U-spin}}}\times \frac{V_{cd}^*V_{us}^{\phantom{*}} 
    \left[V_{td}^* V_{ts}^{\phantom{*}} (f(x_t) - f(x_u)) +
        V_{cd}^* V_{cs}^{\phantom{*}} (f(x_c) - f(x_u))  \right]
    }{\lambda_{sd}} \\ &\sim  \lambda^2 \frac{E_{\overline{K}\overline{K}}}{\mathcal{A}_{K\overline{K}}}\,, \nn
\end{align}
and
\begin{align}
    r_{KK}^C=\frac{C_{KK}}{\mathcal{A}_{K\overline{K}}} &\sim G_F m_D^2\times \frac{(V_{cd}^*V_{us}^{\phantom{*}} V_{ud}^* V_{us}^{\phantom{*}})}{\lambda_{sd}}\times\frac{1}{\varepsilon_{\text{U-spin}}} \sim\lambda^2\frac{C_{\overline{K}\overline{K}}}{\mathcal{A}_{K\overline{K}}} \;.
\end{align}
As a consequence, we find that
\begin{equation}\label{eq:rKK_estimate} 
\left|\frac{r_{KK}}{r_{\overline{K}\overline{K}}}\right|\sim \lambda^2 \sim 0.05 \ll 1.
\end{equation}
We conclude that, due to CKM suppression, at the second order in the weak interaction $D^0\rightarrow \overline{K}^0\overline{K}^0$ gives the dominant contribution to the $|\Delta S|=2$ amplitudes. 

We estimate the relative weak phases of the two relevant diagrams $E_{\Kbar\Kbar}$ and $C_{\Kbar\Kbar}$ as
\begin{align}
\phi_r^{E} &\sim \mathrm{arg}\left(\frac{V_{cs}^* V_{ud}^{\phantom{*}} V_{ts}^* V_{td}^{\phantom{*}} }{\lambda_{sd}}\right) = \mathrm{arg}(\rho + i\, \eta) =  \gamma \sim 1\,, \label{eq:weak-phases-1}\\
\phi_r^{C} &\sim \mathrm{arg}\left(\frac{V_{cs}^* V_{ud}^{\phantom{*}} V_{us}^* V_{ud}^{\phantom{*}} }{\lambda_{sd}}\right) \sim \frac{1}{2}\, A^2\,\eta\, \lambda^4 \sim 3\cdot 10^{-4}\,, \label{eq:weak-phases-2}
\end{align}
where $\lambda$, $A$, $\rho$, and $\eta$ are the Wolfenstein parameters. The necessary higher order contributions to the Wolfenstein expansion can be found,~\emph{e.g.},~in Ref.~\cite{Hocker:2001xe}. We also define $\delta_r^E$ and $\delta_r^C$ to be the relative strong phases of $E_{\Kbar\Kbar}$ and $C_{\Kbar\Kbar}$.
We find that $|r_{\overline{K}\overline{K}}^C|\sim 10^{-4}$ is much larger than $|r_{\overline{K}\overline{K}}^E|\sim10^{-8}$. 
As a consequence of the inverse hierarchy of the magnitude and weak phases,
the two diagrams $E_{\Kbar\Kbar}$ and $C_{\Kbar\Kbar}$ lead to two different CP violating effects in $D^0\rightarrow K_SK_S$, and we keep them both in what follows.

\section{The kaon induced CP asymmetry}\label{sec:CP_asymmetry}

To estimate the kaon induced CP asymmetry, we need to take into account the correlations between the two final state kaons.

We start by considering only the dominant decay amplitude $D^0\rightarrow K^0\overline{K}^0$, that is, we set $r_{\overline{K}\overline{K}}=r_{KK}=0$. 
When two neutral kaons are produced in a two-body decay, they are in a correlated state, and their subsequent decays are not independent \cite{Day_PHI:1961,Lipkin_PHI:1968}. This correlation makes the neutral kaon pair be in an EPR-type entangled state \cite{Einstein_EPR:1935}. In particular, the final state of correlated neutral kaons can be written as:
\begin{equation}\label{eq:KK_state}
    \frac{1}{\sqrt{2}}\left( \ket{K^0(\vec{k})}\ket{\overline{K}^0(-\vec{k})} + (-1)^L \ket{\overline{K}^0(\vec{k})}\ket{K^0(-\vec{k})} \right)\;,
\end{equation}
where, in the $D^0$ rest frame, the two kaons are emitted back-to-back with momenta $\vec{k}$ and $-\vec{k}$, and $L$ is the orbital angular momentum of the system.  Since the $D^0$ in the initial state is a pseudoscalar, conservation of angular momentum implies that the $K^0\overline{K}^0$ system in the final state is in an $s$-wave, that is $L=0$. As a consequence, using Eq.~(\ref{eq:pq}), we find that the correlated state can be written as:
\begin{equation}\label{eq:K_Kbar_final_correlated_state}
    \frac{1}{2\sqrt{2}\;pq}\left( \ket{K_S(\vec{k})}\ket{K_S(-\vec{k})} - \ket{K_L(\vec{k})}\ket{K_L(-\vec{k})} \right)\;,
\end{equation}
where the minus sign comes from the relative sign difference in Eq.~(\ref{eq:pq}). 

We now consider the subleading amplitudes, see Eq.~(\ref{eq:Amplitueds}). In this case, we have two identical kaons in the final state. Bose symmetry implies that we only have the symmetric correlated state, which is written as a simple direct product
\begin{align}\label{eq:KK-iden_state}
\ket{K^0(\vec{k})}\ket{K^0(-\vec{k})}\;,\qquad
\ket{\overline{K}^0(\vec{k})}\ket{\overline{K}^0(-\vec{k})}\;,
\end{align}
respectively. 

Quantum interference and CP violation in correlated states of neutral kaons was studied in the past for the strong decay $\phi\rightarrow K^0\overline{K}^0$~\cite{KLOE-2:2021ila,Buchanan_PHI:1991,Branco:1999fs,Day_PHI:1961,Lipkin_PHI:1968}, which is of particular interest to study decoherence and probe $CPT$ symmetry violation~\cite{Bernabeu:2012nu,Bernabeu:2015aga,KLOE:2006iuj}. From the conservation of the angular momentum, in this channel the $K^0\overline{K}^0$ pair is produced in an antisymmetric $p$-wave state, and the final states with two identical kaons $K^0K^0$ and $\overline{K}^0\overline{K}^0$ are not allowed due to Bose statistics~\cite{KLOE-2:2021ila}.

Here, we adopt the same kind of formalism to determine the total decay rate for the process $D^0\rightarrow K_S(\rightarrow\pi\pi) K_S(\rightarrow\pi\pi)$ parametrized by the decay times $t_1$ and $t_2$ of the two neutral kaons. In particular, we assume that the kaon with momentum $\vec{k}$ decays at a time $t_1$ and the kaon with momentum $-\vec{k}$ at a time $t_2$. By summing coherently over the amplitudes in Eq.~(\ref{eq:Amplitueds}), we find:
\begin{align}\label{Amplitude_tot}
&\mathcal{A} \left(D^0\rightarrow K_S(\rightarrow\pi\pi)_{t_1} K_S(\rightarrow\pi\pi)_{t_2}\right)
=
\mathcal{A}(D^0\to K^0\overline{K}^0) \times
\\
& \qquad\bigg[
\frac{1}{\sqrt{2}}\Big(
\mathcal{A}(K^0(t_1)\to \pi\pi)\,\mathcal{A}(\overline{K}^0(t_2)\to \pi\pi)
+
\mathcal{A}(\overline{K}^0(t_1)\to \pi\pi)\,\mathcal{A}(K^0(t_2)\to \pi\pi)
\Big)
\nonumber\\&
\qquad
+ \vert r_{\overline{K}\overline{K}}\vert e^{i (\delta_r + \phi_r )}\, 
\mathcal{A}(\overline{K}^0(t_1)\to \pi\pi)\,\mathcal{A}(\overline{K}^0(t_2)\to \pi\pi)
\nn\\& \qquad + \vert r_{KK}\vert e^{i (\delta_s + \phi_s )}\,
\mathcal{A}(K^0(t_1)\to \pi\pi)\,\mathcal{A}(K^0(t_2)\to \pi\pi)
\bigg]\,, \nonumber
\end{align}
where 
we used Eqs.~(\ref{eq:KK_state}, \ref{eq:KK-iden_state}) to write the correlated kaon states for each amplitude.
As we discuss in Section~\ref{sec:second_order}, we find that $|r_{\overline{K}\overline{K}}|\sim10^{-4}$ and that $|r_{\overline{K}\overline{K}}|\gg |r_{KK}|$, see Eq.~(\ref{eq:rKK_estimate}). 

We now consider the CP conjugate decay. Using Eq.~(\ref{eq:sep-acp}) we work in the limit of no CP violation in the $D$ decay. The leading order amplitude $\overline{D}^0\rightarrow K^0\overline{K}^0$, has the same correlated neutral kaon state as in Eq.~(\ref{eq:KK_state}). The second biggest contribution, of the order $|r_{\overline{K}\overline{K}}|$, now comes from $\mathcal{A}(\overline{D}^0\rightarrow K^0K^0)$, leading to
\begin{align}\label{Amplitude_tot_BAR}
&\mathcal{A}\left(\overline{D}^0\rightarrow K_S(\rightarrow\pi\pi)_{t_1} K_S(\rightarrow\pi\pi)_{t_2}\right)= 
\mathcal{A}(D^0 \to K^0\overline{K}^0)\times\\
    &\qquad \bigg[
    \frac{1}{\sqrt{2}}\Big(
    \mathcal{A}(K^0(t_1)\to\pi\pi)\,\mathcal{A}(\overline{K}^0(t_2)\to\pi\pi)
    +
    \mathcal{A}(\overline{K}^0(t_1)\to\pi\pi)\,\mathcal{A}(K^0(t_2)\to\pi\pi)
    \Big)\nonumber\\ &
    \qquad
    + \vert r_{\overline{K}\overline{K}}\vert e^{i(\delta_r - \phi_r)} \,\mathcal{A}(K^0(t_1)\to\pi\pi)\,\mathcal{A}(K^0(t_2)\to\pi\pi) \nn\\&\qquad
    + \vert r_{KK}\vert e^{i(\delta_s - \phi_s)}\,\mathcal{A}(\overline{K}^0(t_1)\to\pi\pi)\,\mathcal{A}(\overline{K}^0(t_2)\to\pi\pi)
    \bigg]. \nonumber
\end{align}
We can see, that in the limit $r_{\overline{K}\overline{K}}=r_{KK}=0$, kaon mixing does not generate an asymmetry, as the time dependent decay rates, which are symmetric under exchange of $K^0$ and $\overline{K}^0$, cancel out (note that effects from direct kaon CP violation $\sim \epsilon'$, also cancel out). 

In order to calculate the CP asymmetry in Eq.~(\ref{eq:CP_intro}), we need to square Eqs.~(\ref{Amplitude_tot}, \ref{Amplitude_tot_BAR}), and plug in the time-dependent decay amplitudes:
\begin{align}
    \mathcal{A}(K^0(t)\to \pi\pi) &= \frac{A_S}{2p}\left( e^{-im_S t-\frac{\Gamma_S}{2}t} + \epsilon\,e^{-im_L t-\frac{\Gamma_L}{2}t} \right)\;,\\
    %%%
    \mathcal{A}(\overline{K}^0(t)\to \pi\pi) &= \frac{A_S}{2q}\left( e^{-im_S t-\frac{\Gamma_S}{2}t} - \epsilon\,e^{-im_L t-\frac{\Gamma_L}{2}t} \right)\;.
\end{align}
We find that:
\begin{align}\label{eq:A2_D}
    &\left|\mathcal{A}\left(D^0\rightarrow K_S(\rightarrow\pi\pi)_{t_1} K_S(\rightarrow\pi\pi)_{t_2}\right)\right|^2 = \left| \mathcal{A}(D^0\to K^0\overline{K}^0) \right|^2 \times \Big|A_S\Big|^4 \times \\& \qquad
    \bigg[ \frac{1}{2} e^{-\Gamma_S (t_1 + t_2)}
     + \frac{1}{\sqrt{2}} \vert r_{\bar{K} \bar{K}}\vert e^{-\Gamma_S (t_1 + t_2)} \cos(\phi_r+\delta_r)  \nn\\ & \qquad
    -\frac{1}{\sqrt{2}} \vert \epsilon\vert \vert r_{\bar{K} \bar{K}}\vert  e^{-\Gamma_S (t_1 + t_2)} \times   \bigg(
        -2 \cos(\phi_{\epsilon}  + \phi_r+\delta_r) \nonumber\\
    & \qquad + 
    e^{-\frac{ \Delta \Gamma}{2} t_1}
     \cos( \phi_{\epsilon}  + \phi_r + \delta_r - \Delta m\, t_1) + 
   e^{-\frac{ \Delta \Gamma}{2} t_2}
     \cos(\phi_{\epsilon} + \phi_r + \delta_r  - \Delta m\, t_2 )
        \bigg)\bigg]\,,\nn
\end{align}
and for the CP conjugate,
\begin{align}\label{eq:A2_Dbar}
    &\left|\mathcal{A}\left(\overline{D}^0\rightarrow K_S(\rightarrow\pi\pi)_{t_1} K_S(\rightarrow\pi\pi)_{t_2}\right)\right|^2 = \left| \mathcal{A}(D^0\to K^0\overline{K}^0) \right|^2 \times \Big|A_S\Big|^4 \times  \\ &
    \bigg[ \frac{1}{2} e^{-\Gamma_S (t_1 + t_2)}
     + \frac{1}{\sqrt{2}} \vert r_{\bar{K} \bar{K}}\vert e^{-\Gamma_S (t_1 + t_2)} \cos(-\phi_r+\delta_r) \nn\\&
    -\frac{1}{\sqrt{2}} \vert \epsilon\vert \vert r_{\bar{K} \bar{K}}\vert  e^{-\Gamma_S (t_1 + t_2)} \times   \bigg(
        +2 \cos(\phi_{\epsilon}  - \phi_r+\delta_r) \nonumber\\
    & - 
    e^{-\frac{ \Delta \Gamma}{2} t_1}
     \cos( \phi_{\epsilon}  - \phi_r + \delta_r - \Delta m\, t_1) - 
   e^{-\frac{ \Delta \Gamma}{2} t_2}
     \cos(\phi_{\epsilon} - \phi_r + \delta_r  - \Delta m\, t_2 )
        \bigg)\bigg]\,,\nn
\end{align}
where $\phi_{\epsilon} = \mathrm{arg}(\epsilon)$. 
Note that in Eqs.~(\ref{eq:A2_D}, \ref{eq:A2_Dbar}) we do not write terms of order $\vert \epsilon\vert^2 \sim \vert \epsilon'\vert \sim 10^{-6}$ explicitly, because they cancel in the CP asymmetry, although these contributions are larger than the ones of order $\vert \epsilon\vert\times \vert r_{\bar{K} \bar{K}}\vert \sim 10^{-7}$.
For convenience, in Eq.~(\ref{eq:A2_D}) and Eq.~(\ref{eq:A2_Dbar}) we choose a phase convention such that (see Eq.~(8.87) in Ref.~\cite{Branco:1999fs})
\begin{align}
p &= \frac{1 + \epsilon}{\sqrt{2( 1 + \vert \epsilon\vert^2)}}\,, \qquad
q = \frac{1 - \epsilon}{\sqrt{2( 1 + \vert \epsilon\vert^2)}}\,.
\end{align}
A phase-convention independent treatment of cascade decays has been given in Ref.~\cite{Amorim:1998pi}.
To derive the CP asymmetry we need to subtract Eq.~(\ref{eq:A2_Dbar}) from (\ref{eq:A2_D}).  
Employing the order of magnitude estimates for the diagrams $E_{\Kbar\Kbar}$ and $C_{\Kbar \Kbar}$  
determined in Sec.~\ref{sec:second_order}, we find that there are two separate sources of kaon induced CP asymmetry. As they are small, we treat them additively,
\begin{align}\label{eq:aCPK}
a_{CP}^{K} &= a_{CP}^{K,\epsilon} + a_{CP}^{K,D}\,.
\end{align}
Below we discuss each of them.

\subsection{CP asymmetry in kaon mixing in $D$ decay, $a_{CP}^{K,D}$}

We first consider the asymmetry from the effect of kaon mixing in the $D$ decay. 
Its origin is in the different weak phase between $D^0\to
K^0\Kbar^0$ and $D^0 \to\Kbar^0\Kbar^0$.
If there were no kaon mixing, these two diagrams would result in different final states and could not interfere. It is kaon mixing that makes these two diagrams interfere.

The effect manifests itself as the different sign of the weak phase $\phi_r$ in the leading-order terms proportional to $|r_{\Kbar\Kbar}|$ in Eq.~(\ref{eq:A2_D}) and Eq.~(\ref{eq:A2_Dbar}). 
We find:
\begin{align}\label{eq:aCP_KD_prediction}
    |a_{CP}^{K,D}| = \left|\sqrt{2}\; \sum_{i=E,C} r_{\Kbar\Kbar}^i \sin(\phi_r^i)\sin(\delta_r^i) \right|\lesssim 10^{-8}\;.
\end{align}
Note that the time integral $\int dt_1 dt_2\;e^{-\Gamma_S(t_1+t_2)}$ cancels in the asymmetry.
Note further that both diagrams $C_{\overline{K}\overline{K}}$ and $E_{\overline{K}\overline{K}}$ contribute to the CP asymmetry at the same order of magnitude due to the inverse hierarchy of their magnitudes and weak phases, see Eqs.~(\ref{eq:rKbarKbarE}, \ref{eq:C_KbarKbar}) and Eqs.~(\ref{eq:weak-phases-1}, \ref{eq:weak-phases-2}).

The asymmetry $a_{CP}^{K,D}$ does not depend on the parameter $\epsilon$, as it should. The reason is that it does not depend on CP violation in the neutral kaon system.
It depends solely on the kaon-mixing interference between the final states $K^0\Kbar^0$ and $\Kbar^0\Kbar^0$ in the final state $K_SK_S$.

\subsection{CP asymmetry in neutral kaon oscillations, $a_{CP}^{K,\epsilon}$}
We now consider the asymmetry due to CP violation in  neutral kaon oscillations in the final state. This effect comes from the terms proportional to $\epsilon$ in Eq.~(\ref{eq:A2_D}) and Eq.~(\ref{eq:A2_Dbar}). As we discuss in Section~\ref{sec:second_order}, we find that  $|r_{\overline{K}\overline{K}}|\approx|r_{\Kbar\Kbar}^C|\sim10^{-4}$ and that $|r_{\overline{K}\overline{K}}|\gg |r_{KK}|$, see Eq.~(\ref{eq:rKK_estimate}). Since $|\epsilon|\gg |r_{\overline{K}\overline{K}}^C|\gg|\epsilon|^2$ we only include effects up to next to leading order in this numerical expansion, that is, we neglect effects of $\mathcal{O}(\epsilon^2)$ and $\mathcal{O}(\vert r_{KK}\vert^2)$.
To leading order, we find the result:
\begin{align}\label{eq:aCP_final}
\left| a_{CP}^{K,\epsilon}(D^0\rightarrow K_SK_S)\right| &= \left| \epsilon  \,r^C_{{\overline{K}\overline{K}}}\, \frac{I_{\mathrm{num}}}{I_{\mathrm{denom}}}\right|\,,
\end{align}
with
\begin{align}
&I_{\mathrm{num}} = \sqrt{2} \int dt_1 dt_2 \; e^{-\Gamma_S (t_1 + t_2)} \times\\& \hspace{2mm}\left[
    2 \cos(\phi_{\epsilon} + \delta_r^{C}) \!-\! 
  e^{-\frac{\Delta \Gamma}{2} t_1}
    \cos(\phi_{\epsilon}  + \delta_r^{C} - \Delta m t_1 ) \!-\! 
  e^{-\frac{\Delta \Gamma}{2}  t_2}
    \cos(\phi_{\epsilon} + \delta_r^{C} - \Delta m t_2 )\right], \nonumber \\
&I_{\mathrm{denom}} =\int dt_1 dt_2\;e^{-\Gamma_S(t_1+t_2)}\,.
\end{align}
We showed in Eq.~(\ref{eq:C_KbarKbar}) that $|r^C_{\overline{K}\overline{K}}|\sim 10^{-4}$, while $\delta_r^{C}$ is unknown as it depends on the strong phases of the amplitudes which define the ratio $r^C_{\overline{K}\overline{K}}$. 
Varying the unknown phase $\delta_r^{C}$ in order to maximize Eq.~(\ref{eq:aCP_final}), we find:
\begin{equation}\label{eq:aCP_K_prediction}
    \left|a_{CP}^{K,\epsilon}(D^0\to K_S K_S)\right|\lesssim 10^{-6}\;.
\end{equation}

\subsection{The total asymmetry}
Using Eq.~(\ref{eq:aCP_KD_prediction}) and Eq.~(\ref{eq:aCP_K_prediction}), we conclude that the total asymmetry due to kaon mixing is given by
\begin{equation}\label{eq:aCP_prediction}
    \left|a_{CP}^{K}(D^0\to K_S K_S)\right|= \left| a_{CP}^{K,D}(D^0\to K_S K_S) + a_{CP}^{K,\epsilon}(D^0\to K_S K_S) \right|\lesssim 10^{-6}\;.
\end{equation}
We do not assign an uncertainty to the theory prediction in Eq.~(\ref{eq:aCP_prediction}), as it was derived using an order of magnitude estimate on the amplitudes for $D^0\to K^0\overline{K}^0$ and $D^0\to \overline{K}^0\overline{K}^0$, see Section~\ref{sec:second_order}. Yet, we conclude that this effect is much smaller than the sensitivity of any current realistic experiment, and we expect corrections to this result to be of~$\mathcal{O}(1)$.

We conclude the section with a remark about the importance of the quantum correlation for any general two-body decay into two neutral kaons. When we only consider the leading $\Delta S=0$ decay amplitude into a $K^0\overline{K}^0$ final state, the CP asymmetry in kaon mixing is always zero, regardless of the $(-1)^L$ parity of the correlated state in Eq.~(\ref{eq:KK_state}). In fact, the total time dependent decay rate into the final state of four pions is always symmetric under the exchange of $K^0$ and $\overline{K}^0$, making the CP asymmetry in kaon oscillations vanishing.

We also note that for a correlated neutral kaon pair produced in an odd angular momentum state, such as a $p$-wave in $\phi \to K_S K_L$, there is no CP violation from kaon mixing, even at higher order in the weak interactions. This is because the decay modes into $K^0 K^0$ and $\overline{K}^0 \overline{K}^0$ are forbidden by Bose statistics.

\section{The efficiency function}
Experimentally, a $K_S$ in the final state is defined as two pions 
with an invariant mass $m_{\pi\pi}\approx m_K$. As discussed in Refs.~\cite{Grossman:2011,Grossman:2025}, the reconstructed rate for a realistic experiment depends on an efficiency function that defines the probability to reconstruct $K_S\rightarrow\pi\pi$ as a function of the neutral kaon rest-frame decay time. 

In the case of $D^0\rightarrow K_SK_S$ there are two neutral kaons detected in the final state, and some modifications have to be made.
We define $f_D(t_1,t_2)$ to be the experimental efficiency to reconstruct two neutral kaons produced by a $D^0$ decaying at the times $t_1$ and $t_2$. Equivalently, we define $f_{\overline{D}}(t_1,t_2)$ when the kaons are produced by a $\overline{D}^0$. We then obtain that the reconstructed rates are
\begin{align}
\label{eq:Gamma_rec}
&\Gamma_{\text{rec}}(D^0\rightarrow K_SK _S) = \int dt_1 dt_2\;f_D(t_1,t_2) \Gamma(D^0\rightarrow K_S(\rightarrow\pi\pi)_{t_1} K_S(\rightarrow\pi\pi)_{t_2})\;, \nonumber \\
&\Gamma_{\text{rec}}(\overline {D}^0\rightarrow K_SK _S) = \int dt_1 dt_2\;f_{\overline{D}}(t_1,t_2) \Gamma(\overline{D}^0\rightarrow K_S(\rightarrow\pi\pi)_{t_1} K_S(\rightarrow\pi\pi)_{t_2})\;,
\end{align}
and the measured asymmetry is given by:
\begin{equation}\label{eq:CP_exp}
    a_{CP,exp} = \frac{\Gamma_{\text{rec}}(D^0\rightarrow K_SK_S)-\Gamma_{\text{rec}}(\overline{D}^0\rightarrow K_SK_S)}{\Gamma_{\text{rec}}(D^0\rightarrow K_SK_S)+\Gamma_{\text{rec}}(\overline{D}^0\rightarrow K_SK_S)}\;.
\end{equation} 
For realistic experiments, we expect that the difference in the efficiencies $\Delta f(t_1,t_2) = f_D(t_1,t_2)-f_{\overline{D}}(t_1,t_2)$ to reconstruct the decay products of $D^0$ and $\overline{D}^0$, is a very small effect representing the detector asymmetry $a_{CP}^{\text{det}}$ in Eq.~(\ref{eq:sep-acp}). This difference can arise from effects such as different tagging efficiencies or different energy distributions for $D^0$ and $\overline{D}^0$.

We emphasize that since the two neutral kaons are produced with the same energy in a correlated state, both the time-dependent decay rates and the efficiency functions must be symmetric under exchange of the two kaons, that is under exchange of $t_1$ and $t_2$. Hence, the efficiency function serves as a correction to the ideal result, but it cannot lead to a CP asymmetry due to different mixing effects in $K^0$ and $\overline{K}^0$. This is in contrast with the effects of the efficiency function for decays into two neutral kaons with different spectrum, as discussed in Ref.~\cite{Grossman:2025} for the mode $\tau\rightarrow \nu \pi K^0 \overline{K}^0$. In that case, the $K^0$ and $\overline{K}^0$ are emitted with different energies, which leads to different efficiency functions for detecting $K^0$ and $\overline{K}^0$, and which can induce a CP asymmetry due to kaon oscillation.

\section{Discussion and Conclusions}

The CP asymmetry in $D^0 \to K_S K_S$ decays is expected to be relatively large. In this paper, we show that it is also theoretically clean, in the sense that it does not suffer from contamination due to CP violation in the kaon system. This is in contrast to other decay modes with $K_S$ in the final state, where such effects must be taken into account.

We identify the residual contributions arising from processes that are second order in the weak interaction and estimate their size to be well below the expected CP-violating effects in the $D$ system, and far below any foreseeable experimental sensitivity, see Eq.~(\ref{eq:aCP_prediction}).

Our result can also be applied to other decay modes in which the $K^0$ and $\overline{K}^0$ have identical spectra. One such example is $D^0 \to K^{*0} \overline{K}^{*0} \to K_S K_S \pi^0 \pi^0$. 
More generally, we showed that for any neutral kaon pair produced in a correlated state in a two body decay, the contribution to the kaon mixing CP asymmetry from $K^0\overline{K}^0$ is identically vanishing. This statement is true even when a finite reconstruction efficiency of the experiment is taken into account. Moreover, when the neutral kaon pair is produced with an odd number of orbital angular momentum, the $K^0K^0$ and $\overline{K}^0\overline{K}^0$ states are forbidden by the Bose statistics, and the CP violation in kaon mixing is identically zero even at higher orders in the weak interactions.

\begin{acknowledgments}
YG is supported by the NSF grant PHY-2309456. 
S.S.~is supported by the STFC through an Ernest Rutherford Fellowship under reference ST/Z510233/1 and the grant ST/X003167/1.
\end{acknowledgments}

\bibliographystyle{apsrev4-1}
\bibliography{biblio.bib}

\end{document}